\documentclass[10pt]{article}
\usepackage{epsfig,amsmath,amsthm,graphicx,amssymb,overpic}
\usepackage{epsfig,amsmath,graphicx,amssymb,overpic,cite}
\usepackage{listings,diagbox}

\def\be{\begin{equation}}
\def\ee{\end{equation}}
\def\bee{\begin{eqnarray}}
\def\ene{\end{eqnarray}}
\def\bes{\begin{subequations}}
\def\ees{\end{subequations}}

\def\no{\nonumber}

\def\v{\vspace{0.1in}}
\def\d{\displaystyle}

\begin{document}

%%%%%%%%%%%%%%%%%%%%%%%%%%%%%%%%%%%%%%%%%%%%%%%%%%%%%%%%%%%%%%%%%%%%%%%%%%%%%%%%%
\baselineskip=13pt
\renewcommand {\thefootnote}{\dag}
\renewcommand {\thefootnote}{\ddag}
\renewcommand {\thefootnote}{ }

\pagestyle{plain}

\begin{center}
\baselineskip=16pt \leftline{} \vspace{-.3in} {\Large \bf Data-driven rogue waves and parameter discovery in the defocusing NLS
equation with a potential using the PINN deep learning} \\[0.2in]
\end{center}

\begin{center}
Li Wang and Zhenya Yan$^{*}$\footnote{$^{*}${\it Email address}: zyyan@mmrc.iss.ac.cn (Corresponding author)}  \\[0.03in]
{\it Key Laboratory of Mathematics Mechanization, Academy of Mathematics and Systems Science, \\ Chinese Academy of Sciences, Beijing 100190, China \\
 School of Mathematical Sciences, University of Chinese Academy of Sciences, Beijing 100049, China} \\
 %(Date:\,\, \today)
\end{center}

\vspace{0.3in}

{\baselineskip=13pt

%\begin{tabular}{p{16cm}}
% \hline \\
%\end{tabular}

\vspace{-0.28in}
\begin{abstract}
%\vspace{0.1in}
 The physics-informed neural networks (PINNs) can be used to deep learn the nonlinear partial differential equations and other types of physical models. In this paper, we use the multi-layer PINN deep learning method to study the data-driven rogue wave solutions of the defocusing nonlinear Schr\"odinger (NLS) equation with the time-dependent potential by considering several initial conditions such as the rogue wave, Jacobi elliptic cosine function, two-Gaussian function, or three-hyperbolic-secant function, and periodic boundary conditions. Moreover, the multi-layer PINN algorithm can also be used to learn the parameter in the defocusing NLS equation with the time-dependent potential under the sense of the rogue wave solution.  These results will be useful to further discuss the rogue wave solutions of the defocusing NLS equation with a potential in the study of deep learning neural networks.
\vspace{0.1in}
\end{abstract}


\baselineskip=13pt

\section{Introduction}

\quad With the quick development in computational capabilities and big data generation, the distinct machine learning (ML) approaches have been proposed to explore the effective and accurate results in diverse applications, such as the data mining, computer vision, natural language processing, biomedical analysis, product recommendations, optical text/character recognition, speech recognition, handwriting recognition, data assimilation, genomics, traffic prediction, self-driving cars, medical diagnosis, and stock market trading (see, e.g., Refs.~\cite{Krizhevsky, Lake, Alipanahi, Goodfellow,appl-1,appl-2} and references therein). However, with  the challenges of acquiring data from the distinct application fields,  how to excerpt information effectively and accurately under the condition of partial data missing has becomes an urgent and longstanding problem~\cite{Goodfellow,appl-1,appl-2}. Moreover, many results obtaining from the ML techniques may be incorrect due to the few samples and lack of robustness. In fact, it seems unconvincing for us to draw the conclusions relating to physical laws competently by means of the only ML techniques with the input-output pairs of data, especially for problem with high-dimensional conditions~\cite{Goodfellow}. The  deficiency of the prior laws pertaining to the physical systems may be one of the main reasons. Hence, many researchers tried to use the ML algorithms along with
some physical laws to improve the accuracy of the unknown solutions of physical models~\cite{Lagaris2, Psichogios}.

Recently, with the aid of
sufficient information pertaining to the physical systems, the ML techniques were put forward to establish a both data-efficient
and linear physical-informed neural networks (PINNs)~\cite{Raissi1, Raissi2, Owhadi}.
Moreover, the deduced solutions seemed to be efficient from the ML methods utilizing  Gaussian process regression in combination with a linear operator~\cite{Rasmussen}. Furthermore, the linear problem was also extended to the nonlinear ones, which supplied a novel idea
to recognize and deduce physical models efficiently~\cite{Raissi3, Raissi4}. However, both the local linearizations of nonlinear problems
in the time direction and theoretical assumptions for physical systems, %to adapt to ML algorithms,
 especially for the Bayesian features induced from Gaussian process regression, hinder the promotion and accuracy performance of the algorithm.
In contrast, the method tackling with the nonlinear problems~\cite{Raissi5}  can avoid the limitations~\cite{Raissi3, Raissi4}, and used a deep neural network and a PINN to approximate the observation solution $\widehat w(x,t)$ and
 \bee
 f(x,t):=\widehat w_t-{\cal N}[x,t,\widehat w, \widehat w_x, \widehat w_{xx},...],
\ene
respectively, based on the considered nonlinear evolution partial differential equation (PDE):
\bee \no
 w_t={\cal N}[x,t,w,w_x, w_{xx},...],\quad x\in \Omega,\quad t\in [0, T],
 \ene
where the two considered deep neural networks $\widehat w(x,t)$  and $f(x,t)$ share the same parameters (e.g., weights and biases), and
are able to be trained by minimizing the mean squared error (MSE) loss arising from the initial-boundary value conditions related to the hidden function $\widehat w(x,t)$ and PINN $f(x,t)$. The automatic differentiation technique~\cite{Abadi, Baydin} can be used to perform the partial derivatives of $\widehat w(x,t)$ with
respect to the input variables for the sake of avoiding any additional constraints.
% Except for that, a deep neural network (DNN), as a ML approach, was considered as the nonlinear function approximators~\cite{Hornik1, Hornik2, Hornik3, Cybenko, Petersen}.
%The other similar ideas were also applied to the various formations~\cite{Lagaris, Yadav, Sirignano, Han, Bar-Sinai}.
It should be pointed out that the deep learning approach with the PINNs is powerful since the physical laws arising from the nonlinear
evolution PDEs were considered in the PINNs~\cite{Raissi5}. This similar idea has been applied to the diverse models  with more complicated dynamical behaviors such as soliton solutions of nonlinear evolution PDEs inspired by the
effective methods~\cite{Lagaris2,Psichogios,Raissi5,Yadav, Sirignano, Han,Bar-Sinai,Rudy,raissi20}.

It is well-known that the usual defocusing NLS equation is not capable of  admitting the Benjamin-Feir instability (or modulation instability (MI))~\cite{mi1,mi2}, which is now regarded as one main reason of the rogue wave (RW) generation in the nonlinear wave systems. However, recently an intriguing idea was put forward  to find that the defocusing NLS equation with some time-dependent external potential could support the stable RW solution~\cite{Yan1}. Of course, this idea can also be extended to other related linear or nonlinear wave systems in the mathematical and physical fields~\cite{soliton, soliton2}. In this paper, motivated by the similar ideas about the PINNs~\cite{Lagaris2,  Psichogios, Raissi5,Kondor1, Kondor2, Hirn, Mallat}, we would like to make use of the multi-layer PINN  to focus on the RW solutions of the Cauchy problem of the defocusing NLS equation with a time-dependent potential and initial-boundary value conditions~\cite{opbook,opbook2,Yan1}
\bee\label{dnls}
\left\{\begin{array}{l}
i \psi_{t}  = - \dfrac{1}{2} \psi_{xx} + V(x,t) \psi + \left| \psi\right|^{2} \psi,~~~ x \in (-L, L), ~~~ t \in (-T, T),  \v\\
\psi(x,-T)=\phi_0(x), \quad x\in [-L, L], \v\\
\psi(-L, t) =  \psi(L, t), \quad t\in [-T, T],
\end{array} \right.
\ene
where the subscripts  denote the partial derivatives of the complex field $\psi$ with respect to the space $x$ and time $t$, and $V(x, t)$ denotes the spatio-temporal potential in the form~\cite{Yan1}
\bee
V(x, t)=\frac{4 \left(x^2- t^2\right) - 1}{\left(x^2 + t^2 + 0.25 \right)^2} - 2,   \label{po}
\end{eqnarray}
where $V(x,t)\leq 0$ and $V(x,t)\to -2$ as $|x|,|t|\to \infty$. Eq.~(\ref{dnls}) can be written as a variational form
$i\partial\psi/\partial t=\delta\mathcal{H}/\delta\psi^*$, with the generalized Hamiltonian
\bee
 \mathcal{H}=\int_{-\infty}^{+\infty}\left[\frac12|\psi_x|^2+V(x,t)|\psi|^2+\frac{1}2|\psi|^4\right]dx.
 \ene

The rest of the paper is arranged as follows. Section 2 devotes to demonstrate the data-driven RW solutions of system (\ref{dnls})
in the sense of distinct initial conditions such as the rogue wave, Jacobi elliptic cosine function, two-Gaussian function, or three-hyperbolic-secant function, and the periodic boundary conditions. In Sec. 3, we use the PINN to discuss the identification of the parameter
in Eq.~(\ref{dnls}) with the potential (\ref{po}). Finally, we give the conclusions and discussions in Sec. 4.

\section{The data-driven rogue wave scheme and application}

In this section, we would like to study the data-driven RW solutions of Eq.~(\ref{dnls}) by means of the PINN deep learning method~\cite{Raissi5}. The PINN involving the physical information $f(x, t)$ pertaining to the defocusing NLS equation with the potential (\ref{po}) is chosen as
\begin{equation}
f(x,t):= i \widehat \psi_t  +  \frac{1}{2} \widehat\psi_{xx} - V(x,t) \widehat\psi - | \widehat\psi |^2\widehat\psi, \label{dfnls-p}
\end{equation}
where $\widehat\psi(x,t)=\widehat\psi(x,t; W, b)$  with weights $W$ and biases $b$ denotes the training latent solution generated by utilizing the PINN. Since the two NNs $\widehat\psi(x,t)$ and $f(x,t)$ are both complex-valued functions, thus one can rewrite them as $\widehat\psi(x,t) = \widehat u(x,t) + i\, \widehat v(x,t)$ and $f(x,t)=f_{R}(x,t)+if_{I}(x,t)$, where both  $\widehat u,\,\widehat v,\, f_{R}(x,t)$ and $f_{I}(x,t)$
are all real-valued functions, and $f_R$ and $f_I$ satisfy
 \bee\label{residual}
 \left\{\begin{array}{l}
 f_R(x,t):= -\widehat v_t +\dfrac12 \widehat u_{xx} - V(x,t) \widehat u -  \widehat u(\widehat u^2 + \widehat v^2), \v\\
 f_I(x,t):=  \widehat u_t + \dfrac12 \widehat v_{xx} - V(x,t) \widehat v-  \widehat v(\widehat u^2 + \widehat v^2).
 \end{array}\right.
  \ene
The expression (\ref{dfnls-p}) provides one with some physical laws in the framework of the defocusing NLS equation with a potential, based on which, we  bulid a neural networks $f(x,t)$ full of physical information. Making use of the automatic differentiation technique \cite{Abadi, Baydin}, the arbitrary-order partial derivatives of hidden solution $\widehat \psi(x, t)$ in the condition of the deep neural network (NN) can be gained  by using the chain rule. Furthermore, the PINN $f(x,t)$ shares the same parameters with the NN $\widehat \psi(x, t)$ even though there exits a variety of activation functions during  the differentiation process in  the nonlinear system (\ref{dfnls-p}).

Therefore, with the aid of L-BFGS optimization approach~\cite{Liu}, the common parameters in the hidden function $\widehat \psi(x,t)=\widehat u(x, t)+i \widehat v(x, t)$ and PINN $f(x,t)=f_{R}(x,t)+if_{I}(x,t)$ can be trained by minimizing the whole MSE loss~\cite{Raissi5}
\begin{equation}
MSE = MSE_{int} +  MSE_{b} + MSE_{f}, \label{mse_rw}
\end{equation}%
where the $MSE_{int},\,  MSE_{b},\, MSE_{f}$ are defined by
\begin{align}
MSE_{int} &= \frac{1}{N_{int}} \sum_{j=1}^{N_{int}} \left|\widehat\psi(x_{-T}^{j}, -T)- \psi(x_{-T}^{j}, -T)\right|^{2} \no \\
   &=\frac{1}{N_{int}} \sum_{j=1}^{N_{int}}\left(\widehat u(x_{-T}^{j}, -T)-\left| u(x_{-T}^{j}, -T)\right|^{2}
       + \left|\widehat v(x_{-T}^{j}, -T)- v(x_{-T}^{j}, -T)\right|^{2}\right),  \label{int_rw}   \\
MSE_{b} &= \frac{1}{N_{b}} \sum_{j=1}^{N_{b}}  \left| \widehat\psi(-L, t_{b}^{j}) - \widehat\psi(L, t_{b}^{j}) \right|^{2} \no \\
&= \frac{1}{N_{b}} \sum_{j=1}^{N_{b}} \left( \left| \widehat u(-L, t_{b}^{j}) - \widehat u(L, t_{b}^{j}) \right|^{2}
+\left| \widehat v(-L, t_{b}^{j}) - \widehat v(L, t_{b}^{j}) \right|^{2} \right),
 \label{b_rw}   \\
MSE_{f} &= \frac{1}{N_{f}} \sum_{j=1}^{N_{f}} \left|  f( x_{f}^{j}, t_{f}^{j}) \right|^{2}
= \frac{1}{N_{f}} \sum_{j=1}^{N_{f}}\left( \left|  f_R( x_{f}^{j}, t_{f}^{j}) \right|^{2}+\left|  f_I( x_{f}^{j}, t_{f}^{j}) \right|^{2}\right), \label{f_rw}
\end{align}%
the observed measurements $\{\widehat\psi(x_{-T}^j, -T)=\widehat u(x_{-T}^j, -T)+i\widehat v(x_{-T}^j, -T) \}_{1}^{N_{int}}$ of the hidden field
$\widehat\psi(x,t)$ is linked with the sampled initial training data $\{x_{-T}^j,\, \psi(x_{-T}^{j}, -T)=u(x_{-T}^{j}, -T)+iv(x_{-T}^{j}, -T)\}_{1}^{N_{int}}$ at time $t=-T$, $\{ \widehat\psi(\pm L, t_{b}^j \}_{1}^{N_b}$ is relevant to the selected boundary training data, and $\{ x_{f}^j, t_{f}^j \}^{N_f}_{1}$ is connected with the marked points for the PINN $f(x,t)$. As a result, for the randomly chosen points, $MSE_{int}$ and $MSE_{b}$ represent the  MSE losses of initial and periodic boundary data, respectively, and $MSE_{f}$ is associated with the MSE loss of the PINN (\ref{dfnls-p}).

In the following subsections, we would like to consider the distinct initial conditions and the periodic boundary conditions to study the
data-driven RW solutions of Eq.~(\ref{dnls}) by using the above-mentioned PINN $f(x,t)$.

\subsection{The rogue-wave initial condition}

It is known that Eq.~(\ref{dnls}) with the time-dependent potential (\ref{po}) admits the exact RW solution~\cite{Yan1}
\begin{equation}
\psi(x, t) = \left[ 1 -  \frac{4\left(1+ 2 i t\right)}{4 \left(x^2 + t^2 \right) + 1}  \right] e^{i t}.\label{rw}
\end{equation}
In what follows, we will consider the initial condition $\psi(x,-T)$ of Eq.~(\ref{dnls}) arising from the RW solution (\ref{rw}). To use the PINN deep learning to investigate the Cauchy problem of Eq.~(\ref{dnls}) with the initial condition given by Eq.~(\ref{rw})
  \begin{equation}
\psi(x, -2) = \left( 1 -  \frac{4-16i}{4x^2 + 17}  \right) e^{-2i},\quad x\in [-2.5 \pi,\,  2.5 \pi], \label{rw0}
\end{equation}
  and the Dirichlet periodic boundary condition $\psi( -2.5 \pi, t) = \psi(2.5 \pi, t)$, we imitate Eq.~(\ref{dnls}) with the time-dependent potential (\ref{po}) by means of the Fourier pseudo-spectral method~\cite{Trefethen,yang} to produce the training data for the PINN. The implementation of pseudo-spectral algorithm is to use the discrete Fourier transform in the spatial interval $x\in [-2.5 \pi,  2.5 \pi]$ with $512$ Fourier modes, and a fourth-order Runge-Kutta time-stepping scheme in the temporal interval $[-2, 2]$ with  time-step $\triangle t =2.4\times 10^{-3}$.

 Based on the sampled initial and Dirichlet periodic boundary data,
 %The total observed initial data $\{x_{-T}^j,\, \widehat\psi(x_{-T}^j, -T)\}_1^{N_{int}}$ in the data-driven PINN arises from the hidden solution
 %$\psi(x,t)$ at $t =-T= -0.6$.  In fact,
 the training data-set used in the 10-layer PINN is comprised of the randomly sampled $N_{int} = 100$ points from the
 initial data $\psi(x, -2)$ given by Eq.~(\ref{rw0}), $N_{b} = 100$ points from the periodic boundary data, and $N_{f} =5,000$ collocation points for the PINN $f(x,t)$ given by Eq.~(\ref{dfnls-p}) within the considered spatio-temporal region of the hidden solution
 $\widehat\psi(x,t)$.  Notice that all the marked data in the NN is chosen randomly by mens of the Latin Hypercube Sampling idea~\cite{Stein}. Moreover, the hidden solution $\widehat\psi(x, t) =\widehat u(x, t)+i \widehat v(x, t)$ can be learned  by means of the $8$-hidden-layer deep PINN along with $80$ neurons per layer, and a hyperbolic tangent activation function by
minimizing the MSE loss given by Eq.~(\ref{mse_rw}).

 %Moreover, it is realized that the small marked points can forecast the unknown solution competently in terms of the neural network containing a plenty of physical learning $f(x, t)$  in $MSE_{f}$ part of the loss function.

 Figure~\ref{nlsp} displays the results of the PINN related to the Cauchy problem of the defocusing NLS equation with the time-dependent potential given by Eqs.~(\ref{dnls}), (\ref{po}), and (\ref{rw0}), as well as the periodic boundary condition. Fig.~\ref{nlsp}(a) depicts the magnitude of the hidden RW solution $\widehat\psi(x, t) = \widehat u(x, t)+i \widehat v(x,t)$, together with the locations of some initial-boundary training data.
   %a fairly accurate anticipated results  from the initial training data
Fig.~\ref{nlsp} (b) exhibits the better matches between the learning solution and one derived from the MATLAB
   at three distinct times $t = -0.64,\, 0.03$, and  $0.34$, where the  $\mathbb{L}_{2}$-norm error of $\psi(x,t)$ is 1.3636e-01. Fig.~\ref{nlsp}(c) displays the three-dimensional (3D) profile of the learning RW solution.

 %  Therefore, the multi-layer neural networks applied in the form of the NLS equation with extrinsic potential depending on both time and space make it possible for us to predict the latent solution accurately with only a small amount of sample points.

\begin{figure}[t]
\begin{center}
\vspace{0.05in} {\scalebox{0.56}[0.56]{\includegraphics{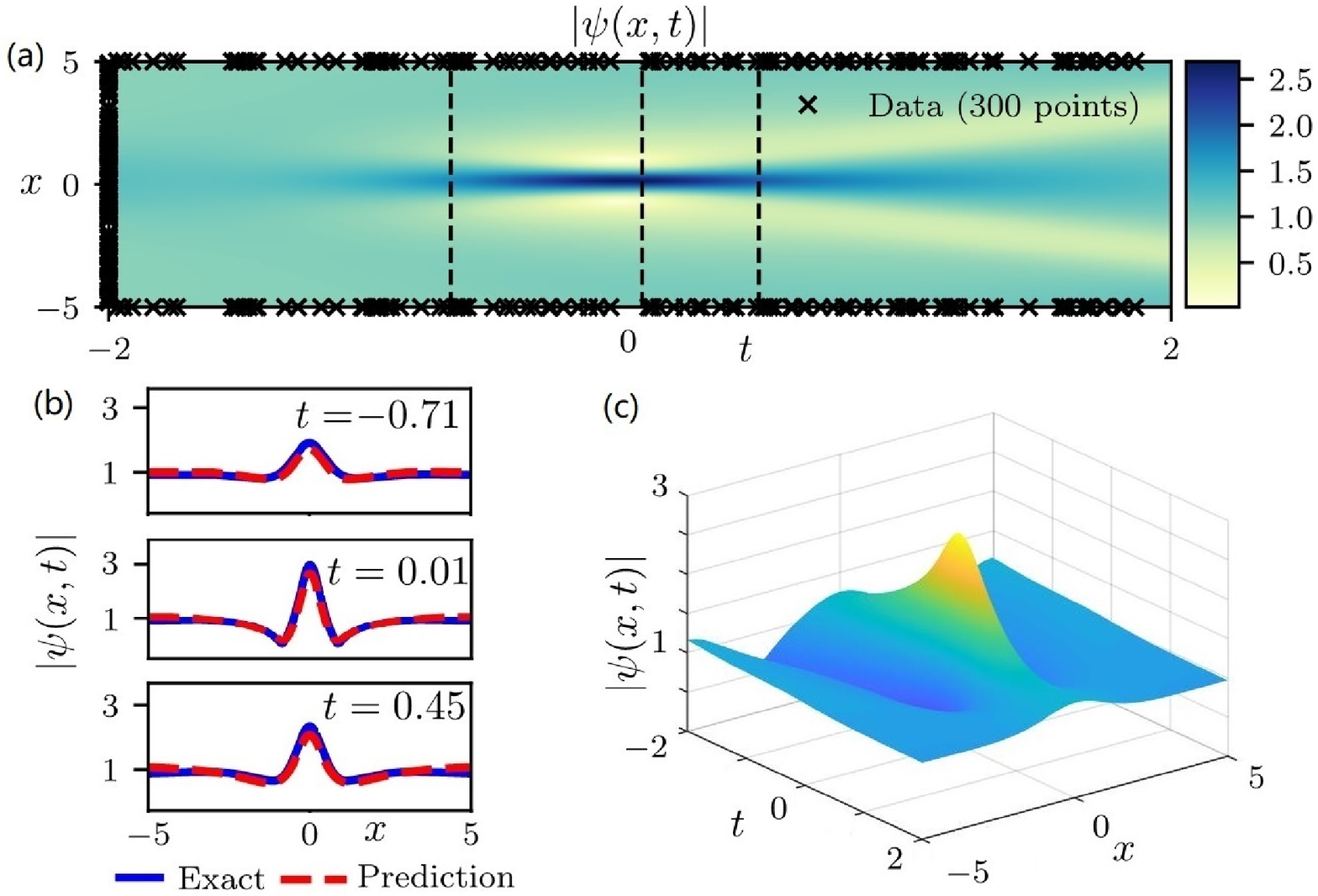}}}
\end{center}
\par
%\vspace{-0.1in}
\caption{{\protect\small (Color online). (a) The data-driven RW solution resulted from the PINN with the randomly chosen $N_{int}=100$ initial points, $N_b=100$ boundary points, and $N_f=5,000$ points for Eq.~(\ref{dfnls-p}) in the hidden solution region, and three distinct tested times $t = -0.75$, $0.01$, and $0.45$ (black dashed lines); (b) The comparisons between the learning and exact RW solutions at the distinct times. The $\mathbb{L}_{2}$-norm error of RW solution $\widehat\psi(x,t)$ is  1.3636e-01; (c) The 3D profile of the learning RW solution.}}
\label{nlsp}
\end{figure}

\begin{figure}[t]
\begin{center}
\vspace{0.05in} {\scalebox{0.56}[0.56]{\includegraphics{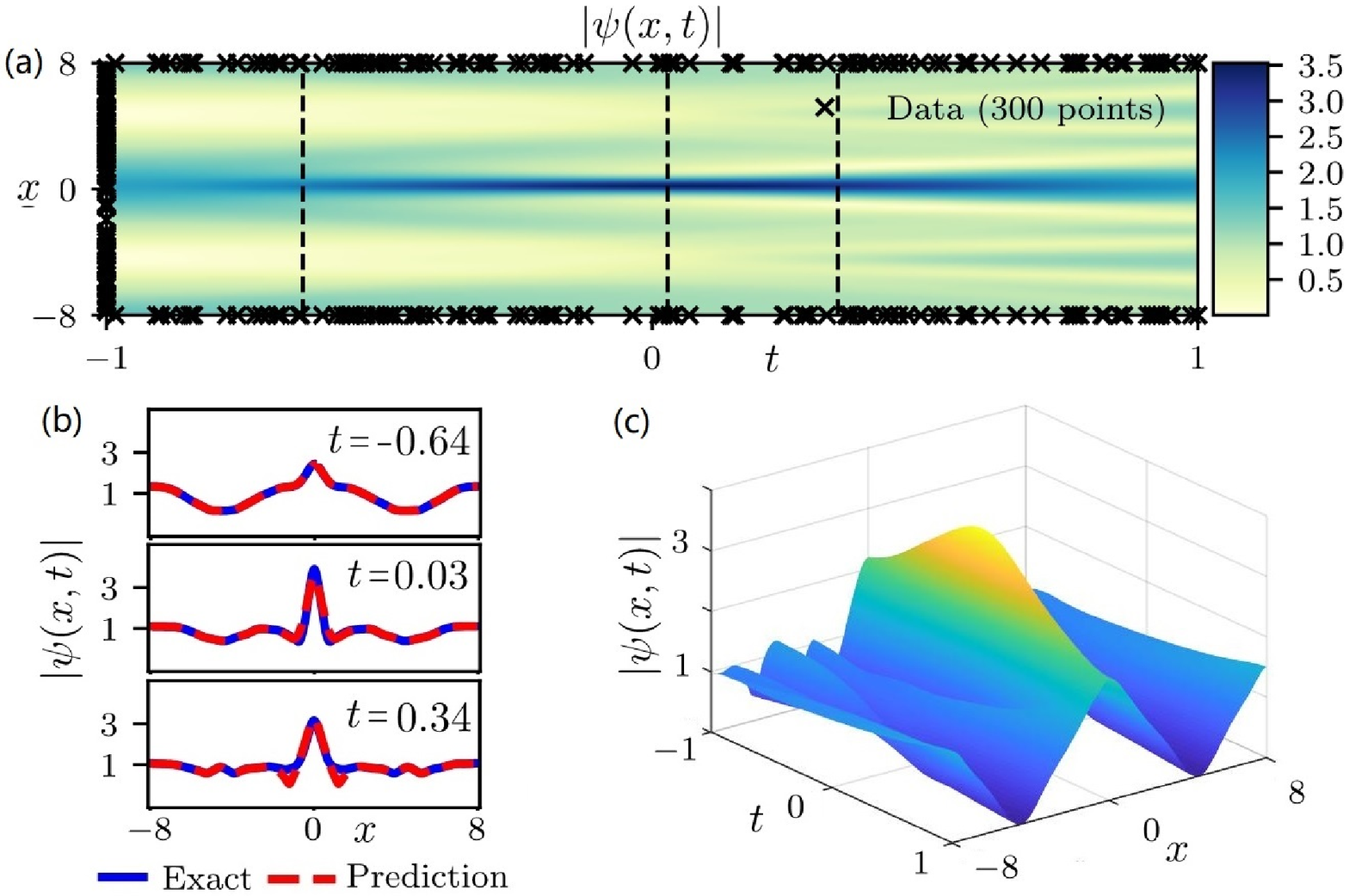}}}
\end{center}
%\par
%\vspace{-0.2in}
\caption{{\protect\small (Color online). (a) The learning RW solution arising from the PINN with the randomly chosen $N_{int}=100$
initial points, $N_b=100$ boundary points, and $N_f=50,000$ points for Eq.~(\ref{dfnls-p}) in the hidden solution region, and the three different  tested times $t = -0.64,\, 0.03$, and  $0.34$ (black dashed lines); (b) The comparison between the learning and exact solutions at the distinct times, where the $\mathbb{L}_{2}$-norm error of the hidden solution is 9.4481e-02; (c) The 3D profile of the learning RW solution.}}
\label{nlsp_2cn}
\end{figure}

\subsection{The double-periodic initial condition}

We here consider another initial condition in the form of the double-periodic function
\begin{equation}
\psi(x, -1) = 1+ {\rm cn}(x, 0.8), \quad x\in [-8, 8], \label{cn}
\end{equation}
and the Direchlet-Neumann periodic boundary conditions
\bee \label{jbc}
\psi(-8,t)=\psi(8, t), \qquad  \psi_x(-8,t)=\psi_x(8, t),  \quad t\in [-1, 1]
\ene
to study the RW generation in the context of Eq.~(\ref{dnls}), where ${\rm cn}(x,m)$ is a Jacobi elliptic cosine function of
modulus $m\in (0,1)$.

By means of the Fourier pseudo-spectral method~\cite{Trefethen,yang} in MATLAB, we exert the double-periodic initial condition (\ref{cn}) and periodic boundary conditions (\ref{jbc}) to acquire the corresponding data for the PINN in the physical framework (\ref{dfnls-p}).  We choose the spatial region
$x\in [-8,  8]$ with the space-step $\triangle x = 0.0312$, and the time region $t\in [-1, 1]$ with time-step $\triangle t=0.004$.
%Obviously, the corresponding conditions for initial input and spatially periodic boundary conditions are selected as $\psi (x,  -1)$  and  $ \psi( -8, t) = \psi(8, t)$, respectively. Furthermore, the observable dynamical behaviours of the preconceived results following  the form of (\ref{dnls}) is a result of the predictable initial solution $\psi (x,  -1)$. And
The sampled points used in the multi-layer neural network are consist of randomly distributed $N_{int} =100$ points from the initial condition $\psi(x, -1)$ given by Eq.~(\ref{cn}), $N_{b} = 100$ points from the periodic boundary conditions (\ref{jbc}),
and $N_{f} =50,000$ points for Eq.~(\ref{dfnls-p}) within the spatio-temporal region of the hidden solution $\widehat\psi(x,t)$. Similarly, by
minimizing the MSE loss given by Eq.~(\ref{mse_rw}) with Eq.~(\ref{b_rw}) replaced by
\bee
\begin{array}{rl}
MSE_{b}&= \d\dfrac{1}{N_{b}} \sum_{j=1}^{N_{b}}\left( \left| \widehat\psi(-L, t_{b}^{j}) - \widehat\psi(L, t_{b}^{j}) \right|^{2}
+ \left| \widehat\psi_x(-L, t_{b}^{j}) - \widehat\psi_x(L, t_{b}^{j}) \right|^{2}\right) \\
&=\d \dfrac{1}{N_{b}} \sum_{j=1}^{N_{b}} \left( \left| \widehat u(-L, t_{b}^{j}) - \widehat u(L, t_{b}^{j}) \right|^{2}
+\left| \widehat v(-L, t_{b}^{j}) - \widehat v(L, t_{b}^{j}) \right|^{2} \right. \\
&\qquad\qquad +\left.\left| \widehat u_x(-L, t_{b}^{j}) - \widehat u_x(L, t_{b}^{j}) \right|^{2}
+\left| \widehat v_x(-L, t_{b}^{j}) - \widehat v_x(L, t_{b}^{j}) \right|^{2} \right),
 \label{b_rwg}
\end{array}
 \ene
  we use the $9$-hidden-layer deep PINN with $80$ neurons per layer, and a hyperbolic tangent activation function to study the Cauchy problem of Eq.~(\ref{dnls}) with Eqs.~(\ref{cn}) and (\ref{jbc}).

%he Latin Hypercube Sampling approach~\cite{Stein} is used to choose the sampled points to train them in a $9$-layer neural network.
%Moreover, the selected randomly spatial-temporal $\{x, t\}$ value contribute to the input of  neural networks containing physical knowledge, meanwhile,  the hidden solution $\psi(x, t) = [  u(x, t), v(x, t) ]$ is taken as output  solution, predicted by the way  of   10-hidden-layer network along with 100 neurons per layer, and a hyperbolic tangent activation function.

 Figure~\ref{nlsp_2cn} displays the deep learning results of the PINN related to the Cauchy problem of the defocusing NLS equation with the time-dependent potential given by Eqs.~(\ref{dnls}), (\ref{po}), (\ref{cn}), and (\ref{jbc}). Fig.~\ref{nlsp_2cn}(a) depicts the magnitude of the
 hidden RW solution $\widehat\psi(x, t) = \widehat u(x, t)+i \widehat v(x,t)$ in the spatio-temporal region $(x,t) \in [-8, 8]\times [-1, 1]$, together with the locations of some initial-boundary training data. Fig.~\ref{nlsp_2cn} (b) exhibits the better matches between the learning solution and one derived from the MATLAB at the three distinct times $t = -0.64,\, -0.37,\,  0.20$, where the $\mathbb{L}_{2}$-norm error of $\widehat\psi(x,t)$ is  9.4481e-02.  Fig.~\ref{nlsp_2cn}(c) exhibits the three-dimensional profile of the learning RW solution.

\subsection{The initial condition consisting of two Gaussian functions and one constant}

We here consider the initial condition consisting of two Gaussian functions and a constant function
\bee
\psi(x,-2) = 1+ e^{-(x-2.5)^2} + e^{(x-2.5)^2}, \quad x\in [-8, 8] \label{rw_2gau}
\ene
and the same periodic boundary conditions (\ref{jbc}), which may also contribute to
the RW generation in the defocusing NLS equation (\ref{dnls}) with the potential (\ref{po}).

Similarly, we exploit the pseudo-spectral method~\cite{Trefethen} in MATLAB to simulate Eqs.~(\ref{dnls}) and (\ref{po}) with the initial-boundary value conditions given by Eqs.~(\ref{rw_2gau}) and (\ref{jbc}), where the spatial region $[-8, 8]$ is separated into $256$ Fourier modes, and
the temporal region $[-2, 2]$ is split into with $501$ steps with time-step $\triangle t = 0.008$. As a result, we can obtain the corresponding
data for the PINN in the physical framework (\ref{dfnls-p}). The training points are made up of the randomly chosen $N_{int} =100$ points from initial data $\psi(x, -2)$, $N_{b} = 100$  points from the boundary data, and $N_{f} =10,000$ points for the PINN $f(x,t)$ given by Eq.~(\ref{dfnls-p}) within the considered spatio-temporal region of the hidden solution $\widehat\psi(x,t)$. Particularly, all the selected collocation points utilized in the PINN are deduced via the Latin Hypercube Sampling method~\cite{Stein}. Therefore, by minimizing the MSE loss given by Eq.~(\ref{mse_rw}) with Eq.~(\ref{b_rw}) replaced by Eq.~(\ref{b_rwg}), we exert a $7$-hidden-layer deep PINN with $60$ neurons per layer, and a hyperbolic tangent activation function to study the initial-boundary problem of Eq.~(\ref{dnls}) with Eqs.~(\ref{rw_2gau}) and (\ref{jbc}).

  Figure~\ref{nlsp_2rw} displays the learning results of the deep PINN related to the Cauchy problem of the defocusing NLS equation with the time-dependent potential given by Eqs.~(\ref{dnls}), (\ref{po}), (\ref{rw_2gau}), and (\ref{jbc}). Fig.~\ref{nlsp_2rw}(a) depicts the magnitude of the
  hidden RW solution $\widehat\psi(x, t) =  \widehat u(x, t)+i\widehat v(x,t)$ in the spatio-temporal region $(x,t)\in [-8, 8]\times [-2, 2]$, in combination with the locations of some initial-boundary training data. Fig.~\ref{nlsp_2rw}(b) exhibits the better matches between the learning solution and one deduced from the MATLAB at the three distinct times $t = -0.51,\, -0.02,\, 0.50$, where the $\mathbb{L}_{2}$-norm error of $\widehat\psi(x,t)$ is  9.6152e-02.  Fig.~\ref{nlsp_2rw}(c) exhibits the three-dimensional profile of the learning RW solution.

\begin{figure}[t]
\begin{center}
\vspace{0.05in} {\scalebox{0.56}[0.56]{\includegraphics{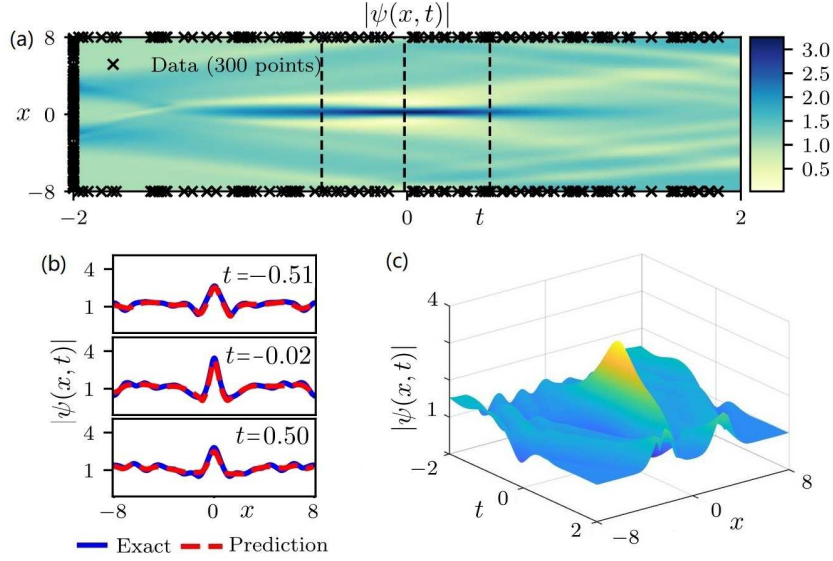}}}
\end{center}
%\par
%\vspace{-0.2in}
\caption{{\protect\small (Color online)  (a) The learning RW solution of the deep PINN with randomly sampled $N_{int}=100$ initial points,  $N_b=100$ boundary points, and $N_f=10,000$ points for the $f(x,t)$ given by Eq.~(\ref{dfnls-p}), and three different  tested  times $t = -0.51, -0.02$, and $0.50$
(black dashed lines); (b) The comparisons between the learning and exact solutions at the distinct times. Moreover, the $\mathbb{L}_{2}$-norm error
of the hidden solution is 9.6152e-02. (c) The 3D profile of the learning RW solution.}}
\label{nlsp_2rw}
\end{figure}

\subsection{The initial condition consisting of three hyperbolic secant functions and one constant}

In this subsection, our aim is to choose the initial condition consisting of three hyperbolic secant functions and one constant described as
 \begin{equation}
\psi(x, -1) = 1 + \sum_{j=-5,0,5}\mathrm{sech}(x+j), \quad x\in [10.5,  10.5], \label{3rw}
\end{equation}
and the Dirichlet-Neumann periodic boundary conditions
\bee \label{jbc-s}
\psi(-10.5, t) = \psi(10.5, t), \quad
\psi_x(-10.5, t) = \psi_x(10.5, t), \quad t\in [-1, 1]
\ene
 to
demonstrate the RW generation in the defocusing NLS equation (\ref{dnls}) with the potential (\ref{po}) via the deep PINN.  We make use of the pseudo-spectral method~\cite{Trefethen} to generate the training data for the PINN. We here consider the spatial region $x\in [10.5,  10.5]$ with $337$ Fourier modes, and temporal domain $t\in [-1, 1]$ with time-step  $\triangle t = 0.004$.

The training data are composed of the randomly sampled $N_{int} =100$ initial points from $\psi(x, -1)$, $N_{b} = 100$ boundary points
from the periodic boundary conditions (\ref{jbc-s}), and $N_{f} =10,000$ collocation points for the PINN $f(x,t)$ given by Eq.~(\ref{dfnls-p})
within the spatio-temporal region of the hidden solution $\widehat\psi(x,t)$. Moreover, all the considered training points
are produced by means of the Latin Hypercube Sampling method~\cite{Stein}. Therefore, by minimizing the MSE loss given by Eq.~(\ref{mse_rw}) with Eq.~(\ref{b_rw}) replaced by Eq.~(\ref{b_rwg}), the deep PINN can be trained to study the RW solution of Eq.~(\ref{dnls}) with the initial-boundary value conditions given by Eqs.~(\ref{3rw}) and (\ref{jbc-s})
by a $8$-hidden-layer PINN with $60$ neurons per layer, and hyperbolic tangent activation function.

Figure~\ref{nlsp_3rw} displays the learning results of the deep PINN related to the Cauchy problem of the defocusing NLS equation with the time-dependent potential given by Eqs.~(\ref{dnls}), (\ref{po}), (\ref{3rw}), and (\ref{jbc-s}). Fig.~\ref{nlsp_3rw}(a) depicts the magnitude of the
  hidden RW solution $\widehat\psi(x, t) =  \widehat u(x, t)+i\widehat v(x,t)$ in the spatio-temporal region $(x,t)\in [10.5,  10.5]\times [-1, 1]$, in combination with the locations of some initial-boundary training data. Fig.~\ref{nlsp_3rw}(b) exhibits the better matches between the learning solution and one deduced from the MATLAB at the three distinct times $t = -0.56,\, 0.04$, and $0.26$, where the $\mathbb{L}_{2}$-norm error of $\widehat\psi(x,t)$ is  1.0832e-02.  Fig.~\ref{nlsp_3rw}(c) exhibits the three-dimensional profile of the learning RW solution. Note that, the neural network in the specific restriction of physical information supplies one way to predict a hidden solution $\widehat\psi(x, t)$ with the highest amplitude $3.636$ under the small sample points.

\begin{figure}[t]
\begin{center}
\vspace{0.05in} {\scalebox{0.56}[0.56]{\includegraphics{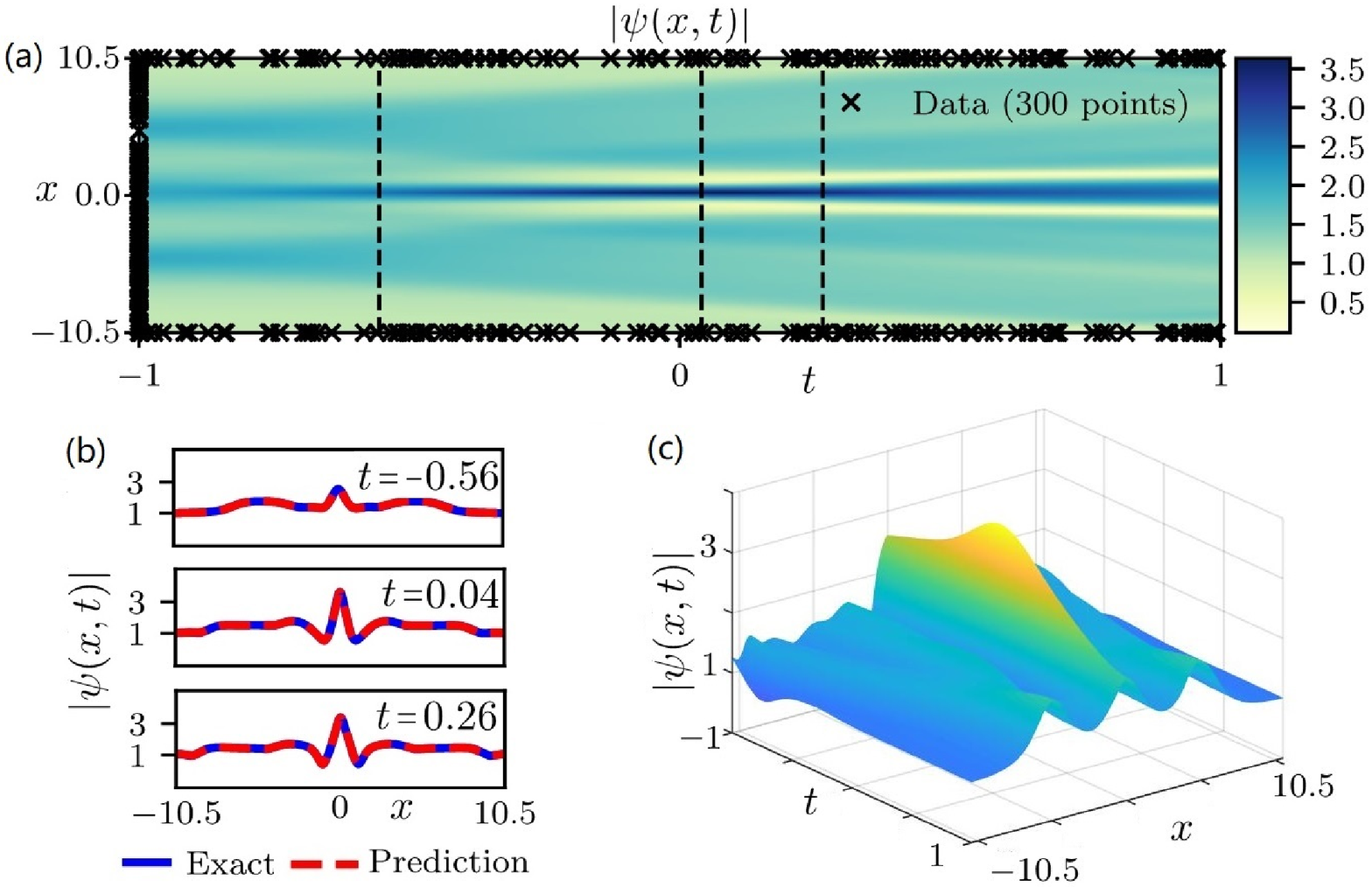}}}
\end{center}
%\par
%\vspace{-0.2in}
\caption{{\protect\small (Color online)  (a) The learning RW solution of the deep PINN with randomly sampled $N_{int}=100$ initial points,  $N_b=100$ boundary points, and $N_f=10,000$ points for the $f(x,t)$ given by Eq.~(\ref{dfnls-p}), and three different  tested  times $t = -0.56,\, 0.04$, and $0.26$
(black dashed lines); (b) The comparisons between the learning and exact solutions at the distinct times. Moreover, the $\mathbb{L}_{2}$-norm error
of the hidden solution is 1.0832e-02. (c) The 3D profile of the learning RW solution.}}
\label{nlsp_3rw}
\end{figure}

\begin{figure}[t]
\begin{center}
\vspace{0.05in} {\scalebox{0.6}[0.6]{\includegraphics{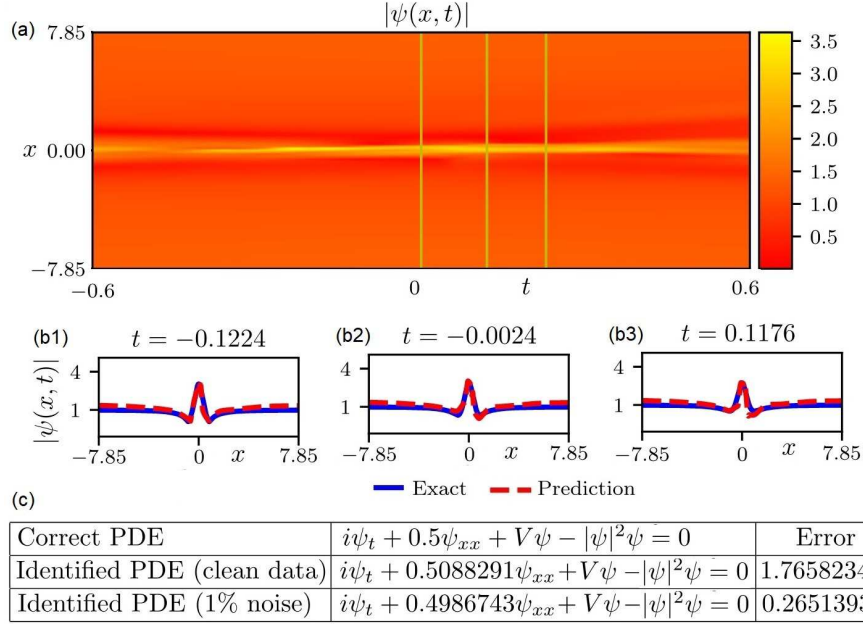}}}
\end{center}
%\par
%\vspace{-0.2in}
\caption{{\protect\small (Color online) (a) The learning RW solution from the PINN with the data in the absence of noise;
(b1)-(b3) The comparison  between the learning and exact RW solutions at the distinct times $t = -0.1224, -0.0024$,  and $0.1176$;
(c) The data-driven parameter ($\lambda$) discovery: correct and identified defocusing NLS equation with the potential.}}
\label{nlsp_iden}
\end{figure}

\section{Data-driven parameter discovery of the physical model}

In this section, we would like to use the PINN to explore the data-driven parameter discovery of the defocusing NLS equation with the time-depdent external potential
\begin{equation}
i \psi_{t}=-\lambda \psi_{xx} + V(x,t) \psi + |\psi|^{2} \psi, \label{nlsp-p}
\end{equation}
where $\psi(x,t)=u(x, t)+iv(x,t)$ ($u,\,v $ are the real and imaginary parts, respectively) is a complex field, and the potential $V(x, t)$ is given by Eq.~(\ref{po}), the coefficient $\lambda$ of the second-order dispersive term  indicates the unknown parameter for training in a multi-layer deep PINN. The complex PINN $f(x,t)$ is defined as
 \begin{equation}
f(x,t):=i \widehat\psi_{t}  + \lambda \widehat\psi_{xx} - V(x,t) \widehat\psi - | \widehat\psi|^{2}\widehat\psi, \label{dnls-f}
\end{equation}
where the hidden function $\widehat\psi(x,t)=\widehat u(x,t)+i\widehat v(x,t)$ with $\widehat u,\, \widehat v $ being the real and imaginary parts, respectively,
and $f(x,t)=f_{u}(x,t)+if_{v}(x,t)$ with $f_u,\, f_v\in \mathbb{R}[x,t]$ satisfying
 \bee
 \left\{\begin{array}{l}
 f_u(x,t):= -\widehat v_t +\lambda \widehat u_{xx} - V(x,t) \widehat u -  \widehat u(\widehat u^2 + \widehat v^2), \v\\
 f_v(x,t):=  \widehat u_t + \lambda \widehat v_{xx} - V(x,t) \widehat v-  \widehat v(\widehat u^2 + \widehat v^2),
 \end{array}\right.
  \ene
which can be learned by training the NN $[\widehat u(x, t), \widehat v(x, t)]$ considered as the outputs. Note that, the parameter $\lambda$ in Eq.~(\ref{nlsp-p}) and  residual construction  $[f_{u}, f_{v}]$ is same as underlying constants in unknown solution $[\widehat u(x, t), \widehat v(x, t)]$, which can be derived in the PINN by minimizing the MSE loss function
\begin{equation}
MSE = \frac{1}{N_p}  \sum_{j=1}^{N_p}\left( \left|\widehat u(x_p^{j},t_p^{j})  - u(x_p^{j},t_p^{j})\right|^{2}
     + \left|\widehat v(x_p^{j},t_p^{j})  -  v(x_p^{j},t_p^{j})\right|^{2}+
      \left|f_u(x_p^{j},t_p^{j})\right|^{2} + \left|f_v(x_p^{j},t_p^{j})\right|^{2} \right).  \label{loss_iden}
\end{equation}

To learn the parameter $\lambda$ in Eq.~(\ref{nlsp-p}) with the aid of the PINN, we need to randomly choose $N_p=1,800$ points from a training data-set for $\psi(x,t)$ in the spatio-temporal region $(x,t)\in [-2.5\pi, 2.5\pi]\times [-0.6, 0.6]$, which is obtained by using the
pseudo-spectral method to simulate the Eq.~(\ref{nlsp-p}) with $\lambda=0.5$ and initial-boundary conditions given by Eq.~(\ref{rw}) with $t=-0.6$ and $\psi(-2.5\pi, t)=\psi(2.5\pi,t)$ in MATLAB. The purpose of pseudo-spectral method is to  employ the discrete Fourier transform to calculate the spatial derives in $x$ direction and to use a fourth-order Runge-Kutta method in the time interval $[-0.6, 0.6]$ with time-step $\Delta t = 0.0024$.
The neural network $[\widehat u(x, t), \widehat v(x, t)]$ and residual network $[f_{u}, f_{v}]$  can be generated from a 9-layer deep PINN with  50 neurons per hidden layer, and the hyperbolic tangent activation function by minimizing the MSE loss (\ref{loss_iden}) via the L-BFGS optimizer~\cite{Liu}.

 Figure~\ref{nlsp_iden}(a) describes the magnitude of the hidden function, $|\widehat \psi(x, t)|= \sqrt{\widehat u^2(x, t) + \widehat v^2(x, t)}$, and three vertical solid lines at the diverse times $t = -0.1224,\,  -0.0024$, and  $0.1776$.  Figs.~\ref{nlsp_iden}(b1)-(b3) display the
 comparisons between the exact and learning solutions at the three distinct times $t = -0.1224, \, -0.0024$ and  $0.1776$, which show
 that the learning and exact solutions match better in the PINN with the small sample points.
   Fig.~\ref{nlsp_iden}(c) exhibits the value of the training parameter $\lambda$ is almost identical to the exact one, where
 the error of $\lambda$ is $1.7658\%$ in the absence of noise, and  $0.265139\%$ in the presence of $1\%$ noise.

\section{Conclusions and discussions}

In conclusion, we have used the multi-layer PINN deep learning method to  study the data-driven rogue wave solutions of the
defocusing NLS equation with the spatio-temporal potential under the distinct initial conditions (e.g., the rogue wave, Jacobi elliptic cosine function, two-Gaussian function, or three-hyperbolic-secant function) and periodic boundary conditions. In particular, when we choose the exact RW solution as the initial condition we find that the PINN can better learn the rogue wave structures. This results imply that the proper initial
condition will be useful to improve the learning results of the PINN. The PINN can also be used to deep learn rogue waves of other nonlinear wave equations in many fields.

Moreover, the data-driven parameter discovery of the defocusing NLS equation with the time-dependent potential under the sense of the rogue wave solution is also studied. It should be pointed out that the considered periodic boundary conditions are non-zero, which differ from the other learning solitons with zero boundary conditions~\cite{Raissi5}, and seem to be difficultly learned. In brief, these results show that the PINNs can be used to learn the rogue waves of the defocusing NLS equation with a spatio-temporal potential even though the small sampled points are applied.
However, there are many unknown issues such as i) there is no the theoretical analysis of the PINNs with different activations, weights and bias functions for indicating latent solutions; ii) what is the more suitable error loss for the different physical models ? iii) whether can the more physical laws make the deep PINNs better learn the corresponding nonlinear physical models ? These problems will be considered in future.

\v \noindent {\bf Acknowledgements} \v

This work is partially supported by the NSFC under Grant Nos. 11731014 and 11925108.

\end{document}